# Voltage tunable sign inversion of magnetoresistance in van der Waals Fe$_3$GeTe$_2$/MoSe$_2$/Fe$_3$GeTe$_2$ tunnel junctions

Shouguo Zhu,[1,2#] Hailong Lin,[1,2#] Wenkai Zhu,[1,2] Weihao Li,[1,2] Jing Zhang,[1] Kaiyou Wang[1,2,*]

[1]State Key Laboratory of Superlattices and Microstructures, Institute of Semiconductors, Chinese Academy of Sciences, Beijing 100083, China
[2]Center of Materials Sciences and Optoelectronics Engineering, University of Chinese Academy of Sciences, Beijing 100049, China

#Equally contributed to this work
*Corresponding author. E-mail: kywang@semi.ac.cn

**Abstracts**

The magnetic tunnel junctions (MTJ) based on van der Waals (vdW) materials possess atomically smooth interfaces with minimal element intermixing. This characteristic ensures that spin polarization is well maintained during transport, leading to the emergence of richer magnetoresistance behaviors. Here, using all 2D vdW MTJs based on magnetic metal Fe$_3$GeTe$_2$ and non-magnetic semiconductor MoSe$_2$, we demonstrate that the magnitude and even sign of the magnetoresistance can be tuned by the applied voltage. The sign inversion of the magnetoresistance is observed in a wide temperature range below the Curie temperature. This tunable magnetoresistance sign may be attributed to the spin polarizations of the tunneling carriers and the band structure of the two ferromagnetic electrodes. Such robust electrical tunability of magnetoresistance extends the functionalities of low-dimensional spintronics and makes it more appealing for next-generation spintronics with all-vdW MTJs.

**Keywords:** Fe$_3$GeTe$_2$, MoSe$_2$, van der Waals ferromagnet, magnetic tunnel junction, sign inversion of magnetoresistance

Ferromagnetic/non-magnetic semiconductor heterostructures form the foundational elements of many spintronic devices, including magnetic tunnel junctions (MTJ)[1,2], spin field-effect transistors[3,4], and spin diode devices[5], etc. These components hold significant promise for the next-generation spintronic technologies. However, the interface spin polarization in traditional three-dimensional (3D) ferromagnetic/non-magnetic heterostructures faces bottlenecks due to difficulties such as lattice mismatch, interface defects, and impedance mismatch[6–8]. In contrast, the novel two-dimensional (2D) van der Waals materials exhibit excellent physical and mechanical properties, such as atomically flat interfaces without dangling bonds[9] and unique spin and valley polarization induced by strong spin-orbit coupling (SOC)[10], providing an ideal platform for improving the interface spin polarization, thereby facilitating the achievement of larger tunneling magnetoresistance (TMR). In heterojunctions combining 3D ferromagnet and 2D semiconductors, researchers have proposed various methods to control spin, such as magnetic proximity effect[11–14], optical spin injection[15], bias-voltage control[16–18], etc. However, the spin polarization at their interfaces and the tunneling magnetoresistance observed in the experiments are still very low; for example, the spin polarizations of 10% (4K)[19] and 4.5% (10K)[20] were observed at the NiFe/graphene and NiFe/MoS$_2$ interfaces, respectively. TMR of 6% (1.4K)[21] and 1.5% (4K)[22] were observed in Co/hBN/Fe and Co/MoSe$_2$/NiFe MTJs, respectively. These relatively low TMR values may relate to the incompatible processes used to deposit and grow 3D ferromagnetic metals on 2D materials, which will inevitably cause interfacial defects and pinholes and lead to a low interface spin polarization.

With the advent of intrinsic 2D magnetic materials[23,24], utilizing 2D ferromagnetic materials such as Fe$_3$GeTe$_2$(FGT), Fe$_3$GaTe$_2$, etc., to construct all-2D van der Waals ferromagnetic/semiconductor heterojunctions aroused great interest. Experiments indicate that devices with linear *I-V* curves tend to show relatively low TMR[25], and further research reveals that this is due to pinholes and defects in the barrier[26]. High-quality vdW heterojunction interfaces without pinholes ultimately solve the problems mentioned above, opening new avenues for enhancing interface spin polarization. A large TMR of 300% was observed in FGT/hBN/FGT MTJ[27], and spin polarizations of 41% (10K) and 70% (10K) were observed at FGT/InSe and FGT/GaSe interfaces[26,28]. More exciting, spin polarization of 55% was achieved at the Fe$_3$GaTe$_2$/WSe$_2$ interface at 300K, resulting in a large room-temperature TMR of 85%[29]. Notably, electrically tunable sign reversal of TMR and spin polarization has been realized in MTJs with hBN[27], WSe$_2$[27], WS$_2$[30], and GaSe[28] barriers. Although experiments have demonstrated enhanced spin polarization in all-2D ferromagnetic/MoSe$_2$ heterojunctions[31], the electrical modulation of spin polarization has remained unexplored yet.

Here, we report the bias and temperature-dependent TMR effect in all 2D Fe$_3$GeTe$_2$/MoSe$_2$/Fe$_3$GeTe$_2$ MTJ devices. We found that the maximum TMR of the device is up to 42% at 10mV @10K, corresponding to a spin polarization of 41.6%, which is ten times higher than that of FGT/MoS$_2$[32] and comparable to FGT/InSe[26] and Fe$_3$GaTe$_2$/MoSe$_2$[31] interfaces. Furthermore, it is important to observe that the TMR sign changes with increasing bias voltage, and a maximum negative TMR of -6% is observed at 10K. The voltage at which this transition occurs remains constant when the ambient temperature changes, confirming that the tunable magnetoresistance behavior is related to the tunneling mechanism. Through qualitative analysis of the electronic structure of FGT, we attribute this to the reversal of net spin polarization of the tunneling electrons at the interface. Our all-2D FGT/MoSe$_2$-based bias-controlled MTJ device provides an effective approach for designing and fabricating low-power and multifunctional spintronic devices such as spin storage and spin logic devices.

The core structure of the tunnel junction is formed by vertically stacking three layers of vdW materials, as shown in Fig. 1a. Both bottom and top layers are FGTs, with different thickness (typically 10-20nm), aiming to realize spin injection and detection. The spacer layer is vdW TMDC semiconductor MoSe$_2$ (~5 nm, Fig. S2), acting as a tunneling barrier to enhance the magnetoresistance. A thick h-BN layer around 30nm is exploited to stack on top of the device, which protects the heterojunctions from air oxidization. Cr/Au electrodes are used to connect the FGT and external two-terminal circuits. A typical device is shown in Fig. 1b, with the blue and green lines representing FGTs and MoSe$_2$, respectively. The overlapping area is controlled under 10 μm$^2$ to prevent the formation of magnetic domains in the tunnel junction[33]. During the magnetotransport measurements, the magnetic field is applied perpendicular to the plane of FGTs. All three MTJs have typical nonlinear *I-V* characteristics, as shown in Fig. 1c, indicating the nature of the tunneling barrier. To further understand the key features of the presence of TMRs, we have summarized some basic parameters of these MTJs in Table. S1.

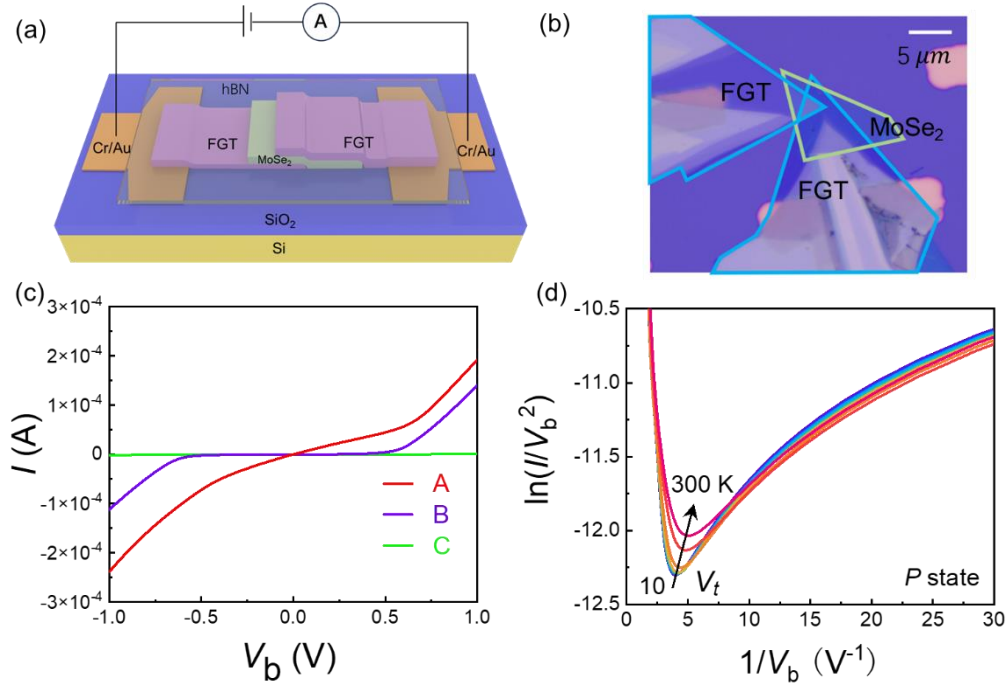

**Fig. 1. Device Structure and *I-V* Characteristics.** (a) The schematic diagram and external experimental configuration of FGT/MoSe$_2$/FGT MTJs. (b) An optical diagram of a typical FGT/MoSe$_2$/FGT MTJ, with the blue and green lines representing the FGT and MoSe$_2$, respectively. Scale bar: 5 μm. (c) *I-V*$_b$ curves of device A - C in bias range of -1.0 to 1.0 V. (d) F-N tunneling relation plot ln(*I*/V$_b^2$) as functions of 1/V$_b$ for the parallel state in device B between 10 – 300 K.

Next, we will mainly discuss the performance of device B, which has nonlinear *I-V* curves (Fig. S3) and relatively large TMR (Fig. S4). At low bias voltages, ln(*I*/V$_b^2$) increases logarithmically with 1/V$_b$, indicating that electrons directly tunnel through the barrier. At higher bias voltages (>0.25 V at 10 K), electrons tunnel into the conduction band of the spacer layer and transport for a certain distance before entering the ferromagnetic detection electrode, named Fowler-Nordheim (F-N) tunneling[34]. At this time, ln(*I*/V$_b^2$) is linearly decreased with 1/V$_b$ (Fig. 1d). The lowest point between direct tunneling and F-N tunneling is the opening voltage (*V$_t$*). The increased thermal kinetic energy of the electrons improves the F-N tunneling probability, resulting in a continuous decrease of *V$_t$*. Similar *I-V* characteristics are also observed in device A (Fig. S5).

The TMR ratio at 10K with different bias voltages ranging from ±0.01V to ±0.5 V was measured by sweeping the perpendicular magnetic field. As depicted in Fig. 2a, the device shows typical spin-valve signals with two stable high (*R$_{AP}$*) and low resistance (*R$_P$*) states. At bias voltage near zero, the highest TMR ratio, defined by $(R_{AP} - R_P)/R_P$, is estimated to be 42%, which is dozens of times higher than that of Co/MoSe$_2$/NiFe MTJs[22]. The different widths of high resistance platforms for antiparallel states at different bias voltages may be related to the unstable pinning field with impurities at the interface. The *I-V* curves of parallel and antiparallel state shown in Fig. 2b can be used to calculate the TMR under continuously varied voltages, which is consistent with the TMR by scanning the magnetic field at different fixed applied voltages, as illustrated in Fig. 2c. It can be seen that the TMR reaches a maximum value at zero bias limit and then decreases with the bias voltage increases. When the positive voltage exceeds 0.34 V (*V$_p$*) and the negative voltage falls below -0.30 V (*V$_n$*), the signs of TMRs switch to negative. Then, the negative TMR increases as the magnitude of the voltage increases, with maximum magnitudes of about -4.7% at 0.53 V and -6.1% at -0.52 V, respectively. After passing the negative peak, the magnitude of the negative TMR begins to decrease and approaches zero. The slight discrepancy in the positive and negative transition voltage is likely due to the imperfectly symmetrical interfaces. The similar bias-dependent TMR behavior is also observed in device A (Fig. S6-7).

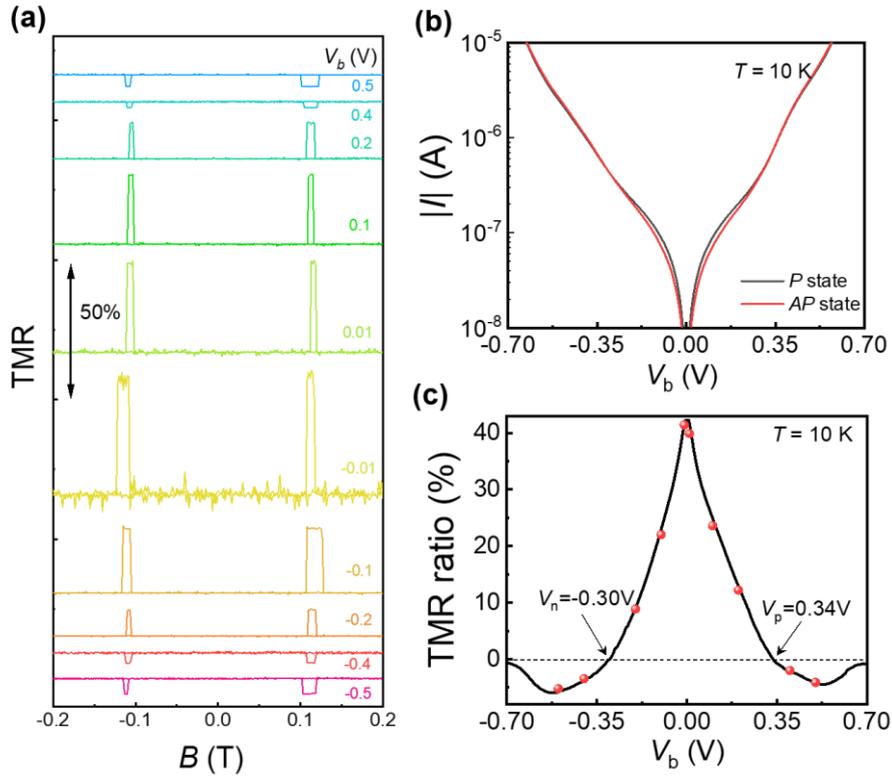

**Fig. 2. TMR in device B.** (a) TMR signals obtained under different bias voltages at 10K. (b) The *I-V* curve for parallel and antiparallel states in logarithmic coordinates. (c) The variation curve of TMR with bias voltage at 10 K. The solid line is calculated from the *I-V* curves of parallel and antiparallel states. The red ball represents the actual value obtained by sweeping the magnetic field.

To systematically evaluate the performance of MTJs, it is necessary to comprehensively obtain the TMR behavior below the Curie temperature ($T_C$) of FGT electrodes. Fig. 3a illustrates the TMR signal at different temperatures under a low bias voltage of 0.01V. The TMR signal decreases as the temperature increases, yet it remains observable at 280K and reaches 5%. As shown in Fig. 3b and Fig. S4, stable sign reversal of TMR in both device B and device A was observed at a wide temperature range below the Curie temperature. Intriguingly, the magnetoresistance behavior in device B remains at 280 K, exceeding the previously reported $T_C$ of the bulk FGT between 220 and 240 K[35]. However, the magnetoresistance can only remain under 220 K in device A (Fig. S6-7). This discrepancy may related to the localized Fe enrichment in FGT flakes leading to an increase in their Curie temperature[36,37]. The specific reasons need to be further explored. Furthermore, it can be seen from Fig. 3b that the magnitudes of $V_p$ and $V_n$ almost have no change with the temperature. This constant $V_p$ and $V_n$ further confirm that the phenomenon of sign inversion is related to the intrinsic band structure of FGT and spin tunneling mechanism[38]. The magnitude of positive TMRs at nearly zero bias decreases approximately as a power function with increasing temperature. The spin polarization at the Femi surface could be calculated by the Julliere model at low bias voltage: *TMR = (2P$^2$/1-P$^2$)*. The maximum spin polarization of 41.6% can be obtained at -0.01 V, which is ten times higher than that of FGT/MoS$_2$[32] and comparable to that of FGT/InSe[26] and Fe$_3$GaTe$_2$/MoS$_2$ devices[31]. The variation of spin polarization with temperature follows a power formula: *P = P$_0$(1-T/T$_C$)$^\beta$*[39], where $P_0$ presents the spin polarization at 0 K, and $\beta$ characterizes the trend of the curve, as illustrated in Fig. 3c.

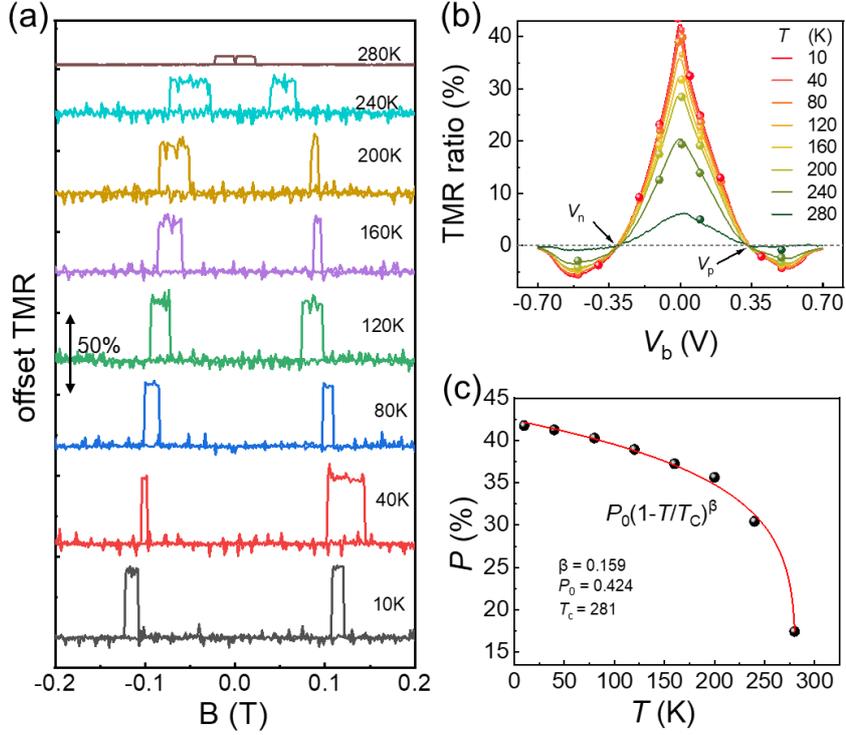

**Fig. 3. Temperature-dependent characteristics of TMR in device B.** (a) The TMR signals under different temperatures at low bias voltage. (b) The variation curves of TMR with bias voltage at different temperatures. Positive ($V_p$) and negative ($V_n$) transition voltages obtained at different temperatures remain unchanged. (c) Spin polarization versus temperature near zero bias calculated by the Julliere model. The red line indicates the fitting result.

The sign reversal of TMR in our devices can be elucidated by a simple model based on the band structure of FGT, which hosts itinerant majority spin-up states near the Fermi level ($E_F$) while maintaining highly correlated electronic bands with spin-down states prevail at about 0.5 eV below $E_F$ [27,40,41], as depicted in Fig. 4. Generally, the conductance of a two-terminal device is related to the spin-resolved density of state (DOS) of the two leads in MTJs. In the parallel configuration, electrons tunnel between the same spin states. The tunneling conductance is proportional to the product of the majority ($D_\uparrow$) and minority ($D_\downarrow$) spin DOS, as expressed by $1/R_P \propto D_\uparrow \times D_\uparrow + D_\downarrow \times D_\downarrow$. In the antiparallel configuration, electrons tunnel between the two spin bands, and the conductance is represented as $1/R_{AP} \propto 2\, D_\uparrow \times D_\downarrow$. When the bias voltage is closed to zero, only the itinerant electrons near the Fermi level participate in tunneling. As shown in Fig. 4a, the DOS of electrons participating in tunneling in the detection electrode is always similar to that of the injection electrode in parallel configuration. Thus, there are enough empty states in the detector to receive the tunneling spins. In the antiparallel configuration shown in Fig. 4b, the majority (minority) spins will tunnel to the empty state of minority (majority) electrons in the detection electrode, thereby reducing the number of conduction channels and leading to higher tunneling resistance. In this case, a positive magnetoresistance is obtained. However, when a finite voltage is applied to the junction, an energy window appears between the injection and detection electrodes, and the expression above should integrate across the energy window. Localized electrons within this bias window have a certain tunneling probability from the injection electrode to the detection electrode. As illustrated in Fig. 4c-d, when the majority spin DOS peak aligns with a valley of majority spin DOS and a peak of empty minority spin DOS in the energy window, the sign reversal of TMR may occur. In this case, the tunneling channel in parallel configuration is less than in antiparallel configuration and thus results in a negative TMR. As mentioned previously, FGT owns an electronic structure consisting of a highly polarized Femi surface and a localized band with reversed spin polarization, thus leading to the sign reversal of TMR under a proper bias voltage. Moreover, there may be more than one transition voltage in FGTs, as confirmed by our observation in device C (Fig. S8). This voltage-dependence of TMR behavior has been confirmed in MTJs with traditional materials and other FGT-based MTJs with h-BN, GaSe, and $WSe_2$ as barrier layers [27,28] (Fig. S9).

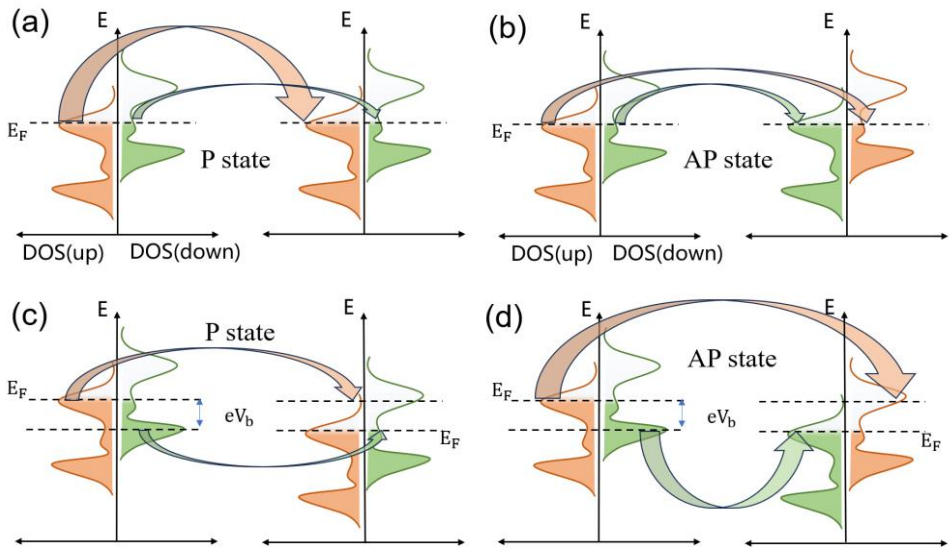

**Fig. 4. The schematic diagram of the energy band for bias voltage tunable TMR sign inversion.** Schematic diagram of the energy band in the (a) parallel and (b) antiparallel configurations when the device is under low bias voltage. Schematic diagram of the energy band in the (c) parallel and (d) antiparallel configurations when the device is under high bias voltage (negative TMR region). The arrows indicate the tunneling channels.

In summary, we have systematically investigated the performance of $FGT/MoSe_2/FGT$ vdW MTJs, where the atomically sharp vdW interface of $MoSe_2$ and FGT can maintain the spin polarization well. The amplitude and even the sign of TMR can be tuned by the bias voltage, attributed to the reversed polarity of high-energy localized electrons in FGT. The dependence of TMR with bias voltage is robust across a wide temperature range below the Curie temperature of FGT. At 10 K, we achieve maximum positive and negative TMR values of 42% and -7%, respectively, with a decline as the temperature increases. Notably, the transition voltage remains constant throughout these temperature variations. Our results indicate that vdW MTJs with flat and sharp interfaces have superior performance compared to their traditional counterparts, and the capability to electrically tune TMR provides an alternative avenue for controlling 2D spintronic devices.

**Experimental details:** A layer of Cr/Au of 10/40 nm on SiO$_2$/Si substrates was prepared by photolithography and Magnetron sputtering. Then, the contact electrodes are formed after a lift-off procedure. FGT and MoSe$_2$ flakes were prepared by mechanical exfoliation. Subsequently, the appropriate thickness and shape of flakes were chosen to stack vdW MTJs using dry transfer methods. This process was aided with the help of an optical microscope and three-axis water hydraulic micromanipulator (MHW-3, Narishige) to realize site control precisely. The whole procedure was performed in a glovebox full of nitrogen with H$_2$O and O$_2$ maintained below 1.0 ppm. Surface morphology and thicknesses of exfoliated flakes were accurately characterized by an atomic force microscope (AFM, MultiMode8, Bruker). The electrical transport properties were investigated using a cryogenic probe station (Lake Shore CRX-VF) with an Agilent B1500A semiconductor parameter analyzer. Additionally, the Raman and photoluminescence spectra of MoSe$_2$ were measured by an optical microscope integrated with a 532 nm wavelength laser (Renishaw inVia-Reflex).

**Supporting information**

Supporting information is available from the

**Data availability**

The data that support the findings of this study are available from the corresponding author upon reasonable request.

**Acknowledgment**

This work was financially supported by the National Key Research and Development Program of China (Grant Nos. 2022YFA1405100), the Beijing Natural Science Foundation Key Program (Grant No. Z220005), the National Natural Science Foundation of China (Grant Nos. 12241405 and 12174384), and the Strategic Priority Research Program of Chinese Academy of Sciences (Grant Nos. XDB44000000 and XDB28000000).

**Conflict of interest**

The authors declare no conflicts of interest.